\documentclass[twocolumn]{jpsj3}
\usepackage{txfonts}
\usepackage{color}

\title{Multiband Superconductivity in KFe$_{2}$As$_{2}$: Evidence for one Isotropic and several Liliputian Energy Gaps}

\author{\name{Fr\'ed\'eric \surname{Hardy}}$^1$\thanks{E-mail: frederic.hardy@kit.edu}, \name{Robert \surname{Eder}}$^1$, \name{Martin \surname{Jackson}}$^2$\thanks{Present address: Department of Low Temperature Physics, Charles University in Prague, 180 00, Praha 8, Czech
Republic}, \name{Dai \surname{Aoki}}$^{3,4}$, \name{Carley
\surname{Paulsen}}$^2$, \name{Thomas \surname{Wolf}}$^1$,
\name{Philipp \surname{Burger}}$^1$, \name{Anna
\surname{B\"ohmer}}$^1$, \name{Peter \surname{Schweiss}}$^1$,
\name{Peter \surname{Adelmann}}$^1$, \name{Robert A.
\surname{Fisher}}$^5$ and \name{Christoph \surname{Meingast}}$^1$}
\inst{$^1$Karlsruher Institut f\"ur Technologie, Institut f\"ur Festk\"orperphysik, 76021 Karlsruhe, Germany \\
$^2$Institut N\'eel, MCBT Department, CNRS and Universit\'e Joseph Fourier, BP 166, 38042 Grenoble cedex 9, France \\
$^3$INAC/SPSMS, CEA Grenoble, 38054 Grenoble cedex 9, France \\
$^4$IMR, Tohoku University, Oarai, Ibaraki 311-1313, Japan \\
$^5$Lawrence Berkeley National Laboratory, Berkeley CA 94720, USA} 

\abst{We report a detailed low-temperature thermodynamic
investigation (heat capacity and magnetization) of the
superconducting state of KFe$_{2}$As$_{2}$ for H $||$ {\it c} axis.
Our measurements reveal that the properties of KFe$_{2}$As$_{2}$ are
dominated by a relatively large nodeless energy gap ($\Delta_{0}$ =
1.9 k$_{B}$T$_{c}$) which excludes $d_{x^{2}-y^{2}}$ symmetry. We
prove the existence of several additional extremely small gaps
($\Delta_{0}$ $<$ 1.0 k$_{B}$T$_{c}$) that have a profound impact on
the low-temperature and low-field behavior, similar to MgB$_{2}$, CeCoIn$_{5}$ and
PrOs$_{4}$Sb$_{12}$. The zero-field heat capacity is analyzed in a
realistic self-consistent 4-band BCS model which qualitatively
reproduces the recent laser ARPES results of Okazaki {\it et al.}
(Science \textbf{337} (2012) 1314). Our results show that extremely
low-temperature measurements, {\it i.e.} T $<$ 0.1 K,will be required in order to resolve the question of the existence of line nodes in this compound.}

\kword{KFe$_{2}$As$_{2}$, iron pnictide, multiband superconductivity, heat capacity, magnetization}

\begin{document}
\maketitle
\section{Introduction}
\begin{figure}[h]
\begin{center}
\includegraphics[width=9cm]{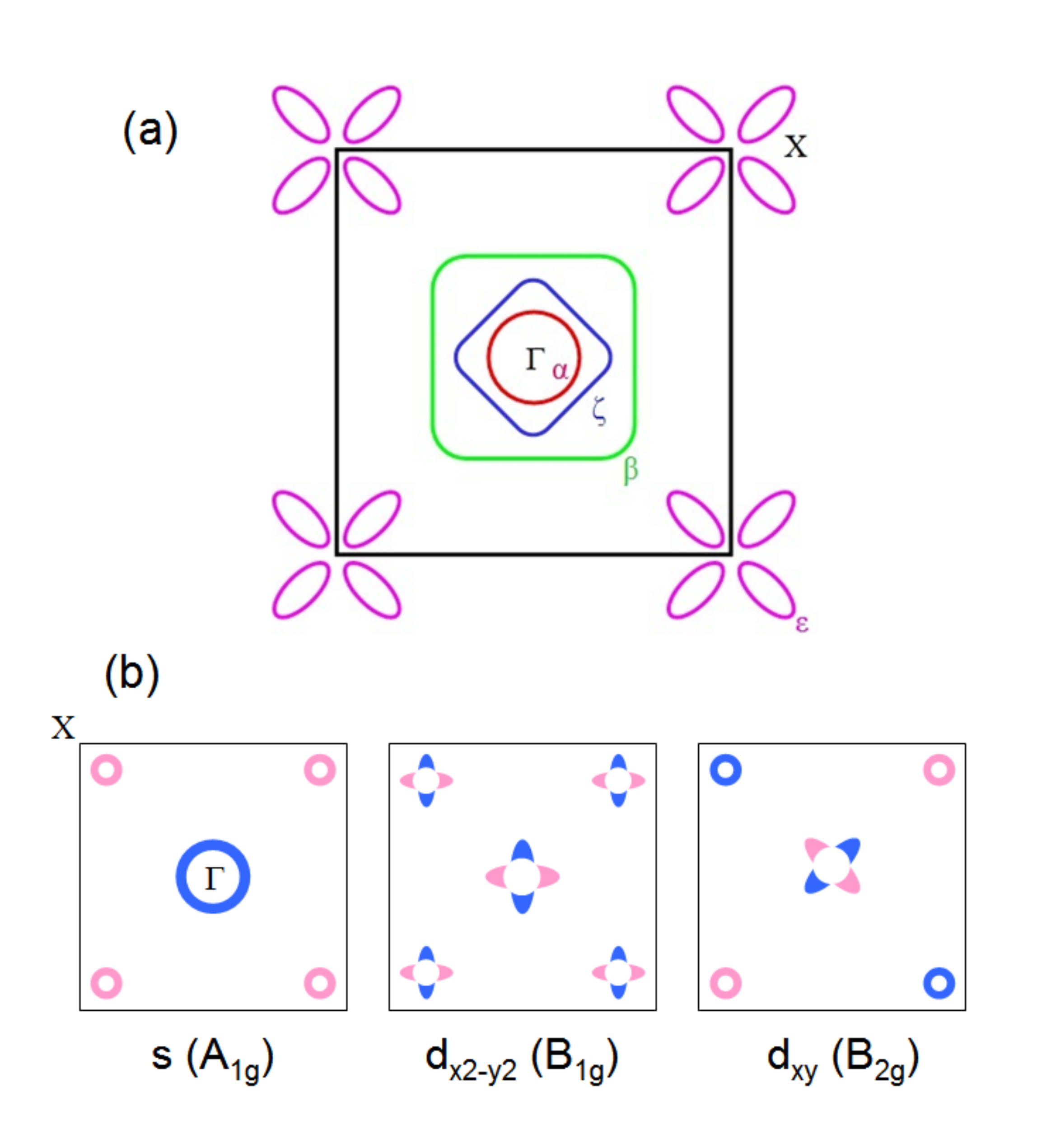}
\caption{\label{fig:Fig0} (Color online) (a) Schematic Fermi surface of KFe$_{2}$As$_{2}$ inferred from dHvA and ARPES measurements.~\cite{Terashima09,Yoshida12} (b) Possible symmetry of the superconducting-state order parameter of KFe$_{2}$As$_{2}$ (only one band is shown at the $\Gamma$ point).}
\end{center}
\end{figure}

The pairing mechanism in iron-pnictide superconductors is still a
subject of intense debate. Similarly to heavy fermions, cuprates and
ruthenates, the proximity of these materials to a magnetic
instability naturally suggests that spin fluctuations can mediate
the formation of Cooper pairs, although other scenarios involving
orbital fluctuations are
possible.~\cite{Mazin08,Kontani10,Chubukov08,Kuroki08,Wang09} In
this context, the symmetry of the superconducting-state order
parameter can have either an $s\pm$ or a
$d$-wave symmetry. Unfortunately, these
states are almost degenerate and the realization of one of these two
states is material-specific, depending on the number and position of
Fermi-surface sheets in the Brillouin zone and their mutual
interactions. In this context, the interpretation of experimental
data is very complicated since the existence (absence) of nodal
behavior does not permit the ruling out of $s$-wave ($d$-wave) symmetry. The Ba$_{1-x}$K$_{x}$Fe$_{2}$As$_{2}$
series is a prominent example. Indeed, at the optimal concentration
($x$ $\approx$ 0.4), heat-capacity~\cite{Popovich10} and
ARPES~\cite{Ding08,Evtushinsky09} measurements give strong evidence
of an $s$-wave state while in the strongly correlated end-member
KFe$_{2}$As$_{2}$, that has only hole pockets (see Fig.
\ref{fig:Fig0}(a)). The situation remains highly controversial.
Recently, thermal-conductivity measurements of Reid {\it et
al.}~\cite{Reid12a} were found to extrapolate at T $\rightarrow$ 0
to a finite residual term $\kappa$(0)/T, independent of sample
purity. This was interpreted as a signature of universal heat
transport, a property of superconductors with symmetry-imposed line
nodes. This hypothetical change from $s$- to $d$-wave symmetry as a
function of doping is allowed theoretically via an intermediate $s$
+ $id$ state that breaks time-reversal
symmetry.~\cite{Lee09,Stanev10,Khodas12,Platt12} On the other hand,
laser ARPES~\cite{Okazaki12} have revealed the existence of
accidental line nodes on only one of the zone-centered pockets in
KFe$_{2}$As$_{2}$, which is only compatible with a nodal
$s\pm$ state.

However, none of these methods are bulk probes of the superconducting state. In this Article, we report a detailed low-temperature thermodynamic investigation (heat capacity and magnetization) of the superconducting state of KFe$_{2}$As$_{2}$. We show quantitatively that the properties of KFe$_{2}$As$_{2}$, including the upper critical field (H$_{c2}$), are dominated by a relatively large nodeless energy gap of amplitude 1.9 k$_{B}$T$_{c}$ which excludes de facto $d_{x^{2}-y^{2}}$ symmetry (see Fig. \ref{fig:Fig0}(b)).
We prove the existence of several additional extremely small gaps ($\Delta_{0}$ $<$ 1.0 k$_{B}$T$_{c}$) and show that they have a profound impact on the low-temperature and low-field behavior, as previously shown experimentally~\cite{Bouquet02,Sologubenko02,Pribulova07a,Seyfarth06,Seyfarth08} and theoretically~\cite{Tewordt03a,Tewordt03b,Barzykin07PRL,Barzykin07,Gorkov12} for MgB$_{2}$, CeCoIn$_{5}$ and PrOs$_{4}$Sb$_{12}$. The zero-field heat capacity is analyzed in a realistic self-consistent 4-band BCS model which qualitatively reproduces the recent laser ARPES results of Okazaki {\it et al.}~\cite{Okazaki12} We also find that extremely low-temperature measurements, {\it i.e.} T $<$ 0.1 K, are required to observe the signature of possible line nodes in KFe$_{2}$As$_{2}$. In accord with recent angle-resolved heat-capacity experiments,~\cite{Kittaka13} our results are compatible with either a $d_{xy}$ or a nodal $s\pm$ state.
\section{Experimental details}
Single crystals of KFe$_{2}$As$_{2}$ were grown in alumina crucibles
using a self-flux method with a molar ratio K:Fe:As=0.3:0.1:0.6. The
crucibles were put and sealed into an iron cylinder filled with
argon gas. After heating up to 700$\,^{\circ}\mathrm{C}$ and then to
980$\,^{\circ}\mathrm{C}$, the furnace was cooled down slowly at a
rate of about 0.5$\,^{\circ}\mathrm{C}$/h. The composition of the
samples was checked by energy-dispersive x-ray analysis and
four-circle diffractometry. The specific heat was measured with a
commercial Quantum Design Physical Property Measurement System
(PPMS) for T $>$ 0.4 K and with a home-made calorimeter for T $<$
0.4 K. For T $>$ 2 K, we used a vibrating sample magnetometer to
measure the magnetization. At lower temperature, magnetization
measurements were performed using a low-temperature superconducting
quantum interference device (SQUID magnetometer) equipped with a
miniature dilution refrigerator developed at the Institut
N\'eel-CNRS Grenoble.~\cite{Burger13}

\section{Zero-field electron specific heat, C$_{e}$ (T, 0)}
\begin{figure}[h]
\begin{center}
\includegraphics[width=9.0cm]{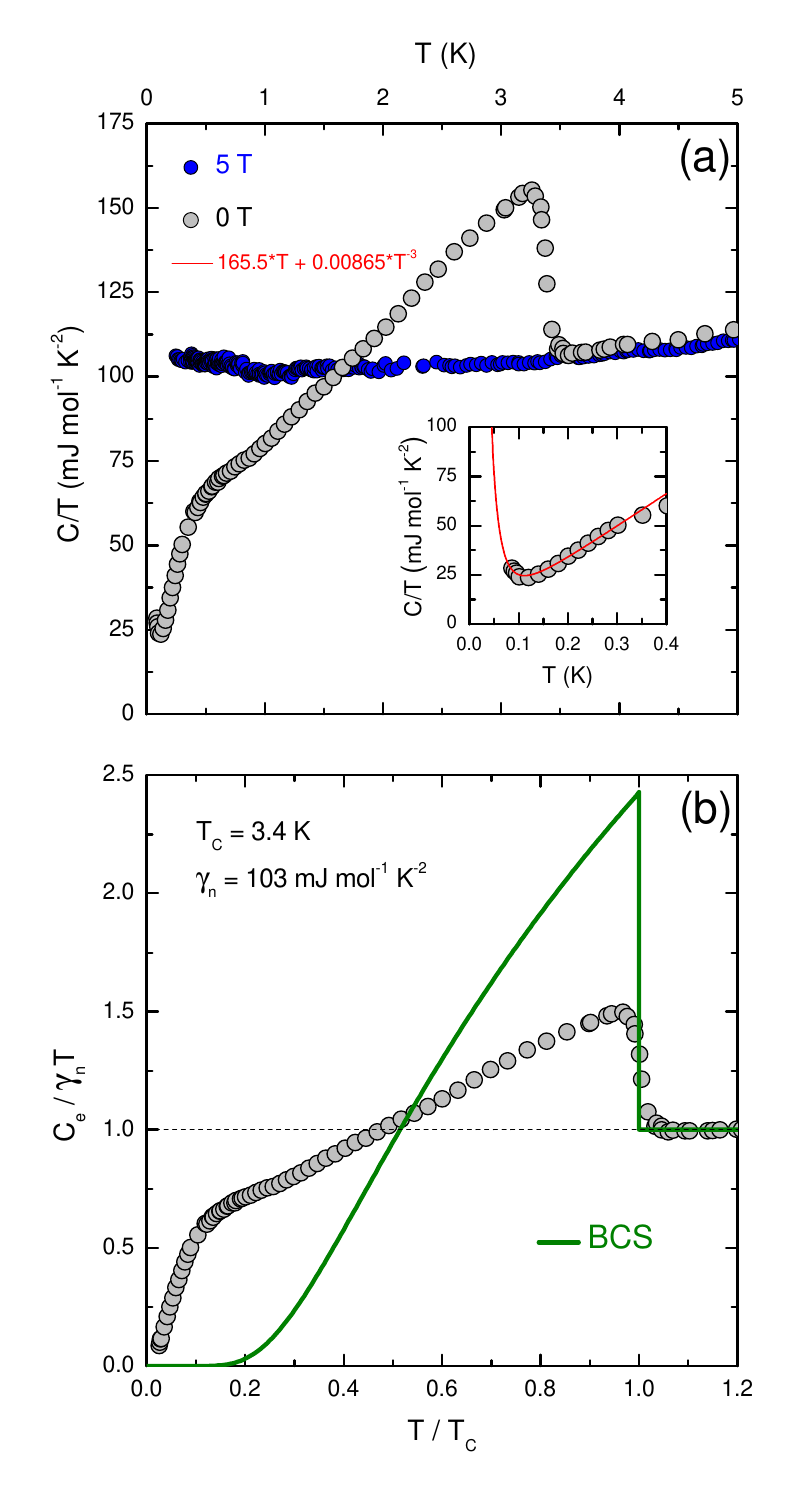}
\caption{\label{fig:Fig1} (Color online) (a) Heat capacity of
KFe$_{2}$As$_{2}$. The inset is a close-up of the
low-temperature region showing the high-temperature tail of a Schottky anomaly. (b) Zero-field electronic heat capacity
C$_{e}$ of KFe$_{2}$As$_{2}$. The green line is the weak-coupling
BCS heat capacity for an $s$-wave superconductor ($\Delta_{0}$=1.764
k$_{B}$T$_{c}$).}
\end{center}
\end{figure}

Figure \ref{fig:Fig1}(a) shows the low-temperature heat capacity of
KFe$_{2}$As$_{2}$. We find a large Sommerfeld coefficient
$\gamma_{n}$ = 103 mJ mol$^{-1}$ K$^{-2}$ and T$_{c}$ = 3.4 K, in
agreement with our previous studies.~\cite{HardyArxiv} Below 0.2K, the high-temperature  tail of a Schottky anomaly, probably due to paramagnetic impurities, is observed (see inset Figure \ref{fig:Fig1}(a)). The electronic contribution C$_{e}$, shown
in \ref{fig:Fig1}(b), is obtained by subtracting, from the measured
data, a Debye term (inferred from the 5 T data), and the Schottky
contribution.

The overall curve bears a strong similarity with
that of MgB$_{2}$.~\cite{Bouquet01} In particular, we observe that:
(i) the jump at T$_{c}$, $\Delta$C/$\gamma_{n}$T$_{c}$ $\approx$
0.54, is substantially smaller than the BCS value ($\Delta$C/C$_{n}$
= 1.43) for a weakly coupled single band $s$-wave superconductor and
(ii) there is a steep quasi-linear decrease of C$_{e}$/T with decreasing
temperature for T/T$_{c}$ $\leq$ 0.1. A similar linear dependence of
the penetration depth was reported by Hashimoto {\it et
al.}~\cite{Hashimoto10} and was interpreted as evidence of line
nodes. In this Article, we argue that this steep feature is instead related to the existence of small energy
gaps ($\Delta_{S}$/k$_{B}$T$_{c}$ $\approx$ T/T$_{c}$ $\approx$
0.2), as inferred from small-angle neutron scattering (SANS)
experiments.~\cite{Kawano11} 

Assuming that all 3 sheets around the
$\Gamma$ point exhibit this tiny gap and using the expression of the
heat-capacity jump of a two-band $s$-wave superconductor in the weak
coupling limit,~\cite{Moskalenko59,Soda66}
\begin{equation}\label{eq:eq1}
\frac{\Delta C}{k_{B}T_{c}}=1.43\cdot\frac{\left(N_{S}\Delta_{S}^{2}+N_{L}\Delta_{L}^{2}\right)^{2}}{\left(N_{S}+N_{L}\right)\left(N_{S}\Delta_{S}^{2}+N_{L}\Delta_{L}^{2}\right)},
\end{equation}
(where the subscripts S and L refer to the small and large gaps,
respectively). We estimate $\Delta_{L}$/k$_{B}$T$_{c}$ $\approx$ 1.8
on the $\epsilon$ band using the individual density of states inferred from dHvA and ARPES
measurements~\cite{Terashima09,Yoshida12} (see Table
\ref{tab:Table1}). Interestingly, we obtain from this simple
estimation a remarkably large gap anisotropy
$\Delta_{L}$/$\Delta_{S}$ $\approx$ 9, in comparison with MgB$_{2}$
where it is about 4.~\cite{Bouquet01}
\begin{table}[h]
\caption{Parameters derived from the dHvA and ARPES measurements assuming 2D Fermi-surface sheets, with $\gamma_{i}$=$\frac{\pi N_{A}k_{B}^{2}a^{2}}{3\hbar^{2}}m_{i}^{*}$ (with $a$ = 3.84 \AA).~\cite{Terashima09,Yoshida12} The last column contains the densities of states used in the 4-band BCS model. m$_{e}$ is the bare electron mass.\\}
\label{tab:Table1}
\begin{center}
\begin{tabular}{c|ccc|c}
              &\multicolumn{3}{c|}{dHvA \& ARPES}&C(T)\\
              \hline
                            &                   m$_{i}^{*}$                     &       $\gamma_{i}$                                &           $\gamma_{i}/\gamma_{n}$     &   N$_{i}$(0)/N(0)                         \\
                            &                   (m$_{e}$)                           &       (mJ mol$^{-1}$K$^{-2}$)     &                                                                           & \\
\hline
$\alpha$            &                   6.06                                    &                   8.8                                     &       0.10                                    &       0.10        \\
$\beta$             &                   17.1                                    &                   24.8                                    &       0.28                                    &       0.31        \\
$\zeta$             &                   11.8                                    &                   17.1                                    &       0.19                                    &       0.23        \\
$\epsilon$      &                   6.62                                    &                   38.4                                    &       0.43                                    &       0.36        \\
\hline
Total                   &                   -                                           &                   90.1                                    &       1.0                                     &       1.0
\end{tabular}
\end{center}
\end{table}
Thus, KFe$_{2}$As$_{2}$ represents a somewhat extreme case of multiband superconductivity similar to the heavy-fermion compounds CeCoIn$_{5}$ and PrOs$_{4}$Sb$_{12}$.~\cite{Seyfarth06,Seyfarth08}
\begin{figure}[t]
\begin{center}
\includegraphics[width=9cm]{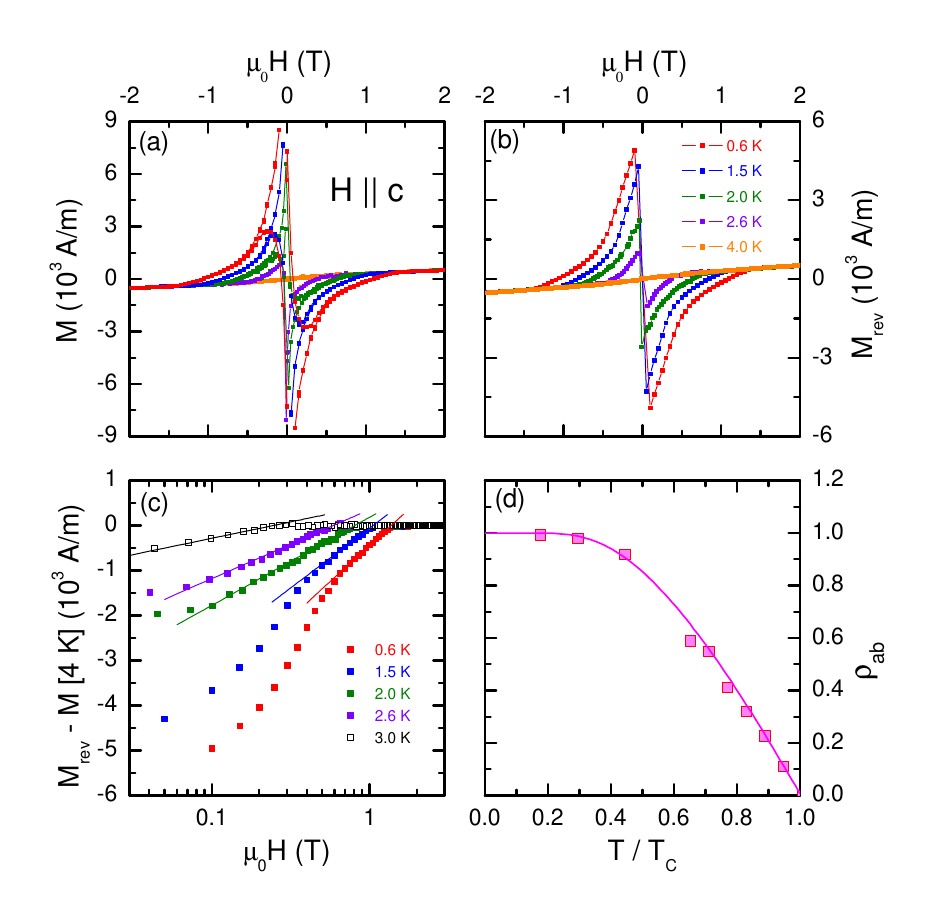}
\caption{\label{fig:Fig2} (Color online) (a) Raw magnetization curves of KFe$_{2}$As$_{2}$ measured at several temperatures for H $||$ $c$. (b) Reversible magnetization curves. (c) Difference of the superconducting- and normal- states reversible magnetizations. (d) In-plane superfluid density derived from the reversible magnetization curves. The solid line is the superfluid density of an $s$-wave gap with $\Delta_{0}$ = 1.9 k$_{B}$T$_{c}$.}
\end{center}
\end{figure}

\section{Mixed-state reversible magnetization, M$_{rev}$ (H)}
Figure \ref{fig:Fig2}(a) shows magnetization curves of
KFe$_{2}$As$_{2}$ for H $||$ $c$ down to 0.6 K. In the normal state
(T = 4 K), a sizeable paramagnetic signal is observed with a
susceptibility of about 3.3 $\times$ 10$^{-4}$ in agreement with our
previous report.~\cite{HardyArxiv} At lower temperatures, the
magnetization curves are reversible over a wide field interval ({\it
e.g.} H$_{c2}$/2.5 $<$ H $<$  H$_{c2}$ = 1.4 T at 0.6 K), indicating
a small concentration of pinning centers as confirmed by the
observation of a well defined hexagonal vortex lattice by
SANS.~\cite{Kawano11} Together with the observation of quantum
oscillations and the absence of significant residual density of
states in the limit T $\rightarrow$ 0 (see Fig. \ref{fig:Fig1}(b))
this indicates that our KFe$_{2}$As$_{2}$ single crystals are weakly
disordered and are in the clean limit. Note that this is at odds with Co-doped
BaFe$_{2}$As$_{2}$ samples in which no vortex lattice could be
observed.~\cite{Eskildsen09,Inosov10}  As a result, accurate reversible
magnetization curves can be obtained for our sample by averaging the increasing and
decreasing branches of the magnetization loop, as illustrated in
Fig. \ref{fig:Fig2}(b). 

In single-band type II superconductors,
M$_{rev}$(H) is entirely defined by H/H$_{c2}$ and the
Ginzburg-Landau parameter $\kappa$ =
$\lambda$/$\xi$,~\cite{Brandt03} with
\begin{equation}\label{eq:eq2a}
M_{rev}=\frac{H-H_{c2}}{(2\kappa^{2}-1)\beta_{A}+1},
\end{equation}
at high field (Abrikosov regime, with $\beta_{A}$ the Abrikosov coefficient). In the intermediate field range (London regime),~\cite{DeGennesBook,Kogan88} the reversible magnetization is linear in the logarithm of the applied field with
\begin{equation}\label{eq:eq2b}
\mu_{0}M_{rev}=-\frac{\phi_{0}}{8\pi \lambda}\ln{\left(b\frac{H_{c2}}{H}\right)},
\end{equation}
where $\lambda$ = $\lambda_{ab}$ is the in-plane penetration depth for H $||$ $c$ and $b$ a constant.~\cite{Lang92}  

Thus, in close analogy to the case of MgB$_{2}$,~\cite{Klein06} Fig. \ref{fig:Fig2}(c) shows that the linear evolution of M$_{rev}$(H) expected near H$_{c2}$ is not observed and the London dependence dominates up to H$_{c2}$,which therefore allows the determination of $\lambda_{ab}$(T) using Eq.(\ref{eq:eq2b}). The derived superfluid density, defined as $\rho_{ab}$(T) = $\left[\lambda_{ab}(\textrm{0.6 K})/\lambda_{ab}(T)\right]^{2}$, is shown in Fig. \ref{fig:Fig2}(d) together with the calculation for an $s$-wave gap of amplitude $\Delta_{0}$ = 1.9 k$_{B}$T$_{c}$, which accurately reproduces the data. As a consequence, our analysis firmly establishes the existence of a relatively large nodeless gap in KFe$_{2}$As$_{2}$, in agreement with our rough above-mentioned heat-capacity analysis. Contrary to direct penetration-depth measurements,~\cite{Hashimoto10} our estimate of $\rho_{ab}$(T) was inferred from high-field data where the large vortex cores related to the smaller gaps have already overlapped, as observed in MgB$_{2}$,~\cite{Eskildsen02} and discussed hereafter. This explains why this indirect derivation of $\rho_{ab}$(T) is only sensitive to the larger gap as found for MgB$_{2}$.~\cite{Zehetmayer04,Zehetmayer13} \\

\section{Mixed-state specific heat, $\gamma$ (H)}
\begin{figure}[h]
\begin{center}
\includegraphics[width=9cm]{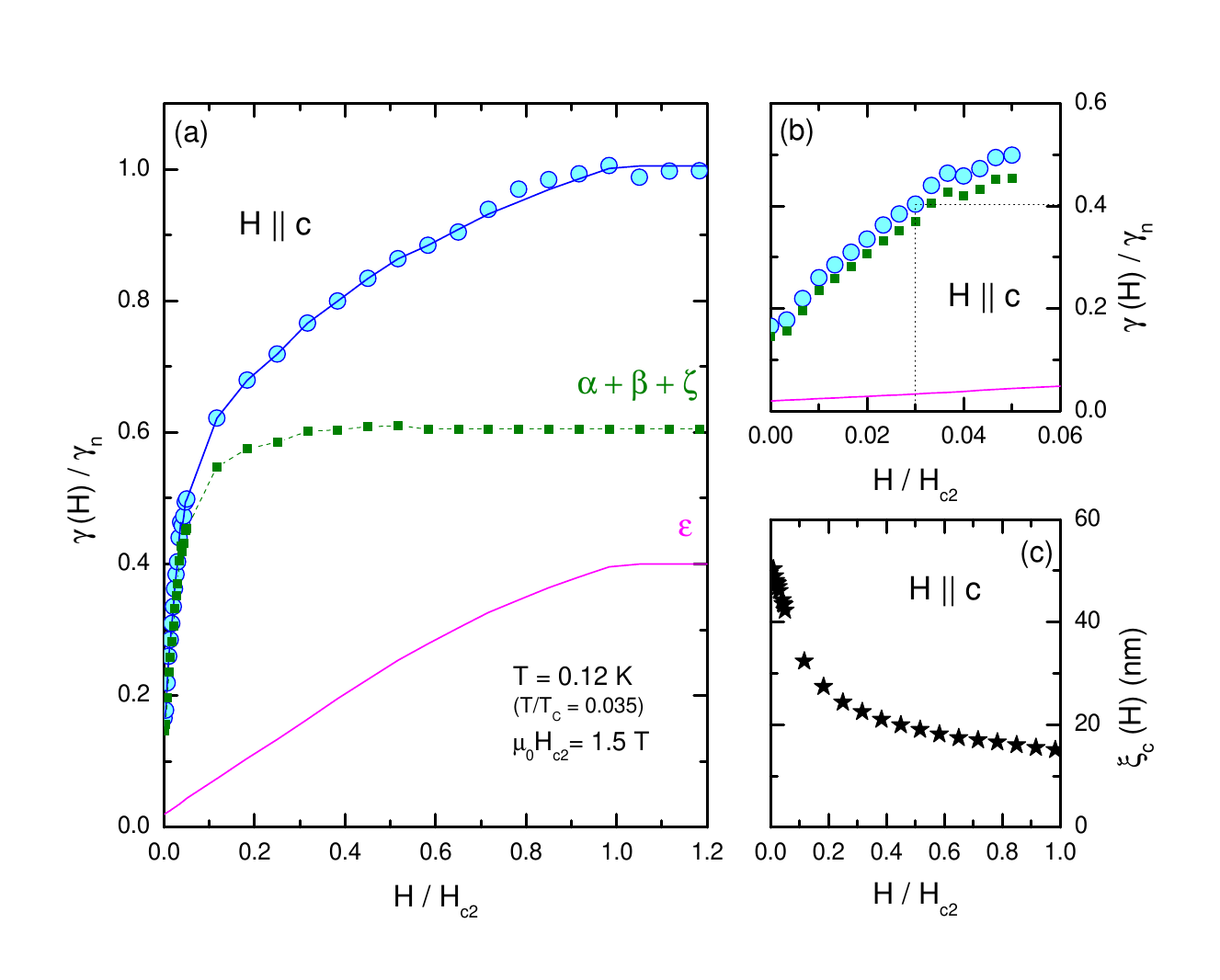}
\caption{\label{fig:Fig3} (Color online) (a) Field dependence of the
heat capacity of KFe$_{2}$As$_{2}$ (blue symbols) at T = 0.12 K for
H $||$ $c$. The magenta line is the theoretical curve of the
mixed-state heat capacity of an $s$-wave
superconductor~\cite{Ichioka07} normalized by the density of states
of the $\epsilon$ band (see Table \ref{tab:Table1}). The green curve
is the resulting contribution of the 3 small energy gaps $\alpha$,
$\beta$ and $\zeta$ obtained by subtracting the heat capacity of the
$\epsilon$ band from the data. (b) Close-up of the low-field region.
(c) Field dependent vortex core size derived from the mixed-state
heat capacity.}
\end{center}
\end{figure}

Evidence for the existence of tiny energy gaps can
be found using heat-capacity measurements in the mixed state as
previously shown for pure, Al- and C-doped
MgB$_{2}$.~\cite{Bouquet02,Pribulova07a,Pribulova07b,Fisher13}
Figure \ref{fig:Fig3}(a) shows the field dependence of the electron
heat capacity $\gamma$(H) at 0.12 K ({\it i.e.} T/T$_{c}$ = 0.035)
for H parallel to $c$. Similar to MgB$_{2}$, we find that
$\gamma$(H) is very non-linear with applied magnetic field. In
very low fields, $\gamma$(H) increases abruptly and reaches
$\gamma$(H)/$\gamma_{n}$ $\approx$ 0.6 at only H/H$_{c2}$ $\approx$
0.1. In larger fields, $\gamma$(H) closely follows the behavior
expected for an individual band (magenta line),~\cite{Ichioka07}
indicating that the interband couplings between the sheet with the
largest gap and the other bands is rather small.~\cite{Tewordt03b}
Using the densities of states derived from dHvA and ARPES (see Table
\ref{tab:Table1}), we can unambiguously ascribe the largest gap to
the $\epsilon$ band and subtract its contribution from $\gamma$(H)
to obtain the mixed-state heat capacity of the remaining $\alpha$,
$\beta$ and $\zeta$ bands (green curve in Figs. \ref{fig:Fig3}(a)
and \ref{fig:Fig3}(b)). We find that these bands have almost
recovered their normal-state value in a 'crossover' field
H$^{S}_{c2}$ $\approx$ 0.1 $\times$ H$_{c2}$ which is defined here
as,
\begin{equation}\label{eq:eq2c}
H^{S}_{c2}\approx H_{c2}\left(\frac{\xi_{S}}{\xi_{L}}\right)^{2},
\end{equation}
and which would correspond to the upper critical field of the small
gaps in the absence of interband couplings.~\cite{Tewordt03b} Thus,
the disappearance of the small gaps associated with the $\alpha$,
$\beta$ and $\zeta$ sheets in an applied field is
more rapid than that of the $\epsilon$ band which shows a
conventional individual dependence. Following Klein {\it et
al.},~\cite{Klein06} we assume that all the excitations are
localized in the vortex cores ({\it i.e.} we neglect the small-gap
Doppler shift~\cite{Bang10}) and that the system can be described by
only one field dependent quantity, $\xi_{c}$(H), which is a
measure of the vortex-core size. In this context, $\gamma$(H)
$\propto$ $\gamma_{n}\cdot \left(\xi_{c}(H)/d\right)^{2}$ (where $d$
$\propto$ 1/$\sqrt{H}$ is the intervortex distance), and we obtain
directly $\xi_{c}$(H) as shown in Fig. \ref{fig:Fig3}(c), with
$\xi_{c}$(H=H$_{c2}$)=$\sqrt{\Phi_{0}/2\pi\mu_{0}H_{c2}}$. We find
that the vortex-core size smoothly decreases from 50 to 15 nm in
high fields, explaining the smooth evolution of the contributions of
the $\alpha$, $\beta$ and $\zeta$ bands to $\gamma$(H) near
H$^{S}_{c2}$. Thus, the small gaps on these sheets remain finite due
to nonzero interband coupling even for H $>>$ H$^{S}_{c2}$ where
their vortex cores overlap. In the opposite limit, {\it i.e.} where
the Doppler shift (Volovik effect) dominates, we note that a similar
dependence of $\gamma$(H) for the small gaps (see Fig.
\ref{fig:Fig3}(b)) is expected theoretically for
$\Delta_{S}/\Delta_{L}$ $\approx$ 0.1.~\cite{Bang10}

\section{Comparison with thermal-conductivity measurements, $\kappa$ (T, H)}
In light of our results, we comment here on the
interpretation of recent heat-transport experiments in
KFe$_{2}$As$_{2}$. In Refs.~\cite{Reid12a} and~\cite{Reid12b},
thermal-conductivity measurements $\kappa$(T)/T, performed for T $>$
0.1 K, were found to extrapolate at T $\rightarrow$ 0 to a finite
residual term $\kappa$(0)/T independent of sample purity. This was
interpreted as a signature of universal heat transport, a property
of superconductors with symmetry-imposed line nodes such as $d$-wave
states. Experimentally, these measurements were not strictly
realized in zero magnetic field because it was necessary to apply a small field of 0.05
T ({\it i.e.} H/H$_{c2}$ $\approx$ 0.03) to suppress
superconductivity of the soldered contacts. However, as shown in
Fig. \ref{fig:Fig3}(b), this small field is large enough to produce
an enhancement of the density of states, reaching 40\% of the normal-state value at 0.12 K, which inexorably leads to
a finite value of $\kappa_{0}$/T. In addition, our specific-heat
measurements show that a significant increase of $\kappa$(H)/T is also to be 
expected for H/H$_{c2}$ $<$ 0.1 due to these small gaps. This
feature was not observed in Refs.~\cite{Reid12a},~\cite{Reid12b}
or in the more recent data of Watanabe {\it et
al.}~\cite{Watanabe13} while it clearly appears in many other
multiband superconductors including MgB$_{2}$, CeCoIn$_{5}$ and
PrOs$_{4}$Sb$_{12}$.~\cite{Sologubenko02,Seyfarth06,Seyfarth08}
Thus, the origin of the finite $\kappa_{0}$/T cannot be attributed
to $d$-wave superconductivity in KFe$_{2}$As$_{2}$ in these
experimental conditions. We note that the use of superconducting
solder was already pointed out to produce spurious results in
Ref.~\cite{Seyfarth08}. On the other hand, our observation of a
relatively large isotropic gap does not rule out definitively $d$-wave
superconductivity in KFe$_{2}$As$_{2}$. Actually, only the
$d_{x^{2}-y^{2}}$ order parameter, with nodes located on the
diagonals of the Brillouin zone, is excluded while $d_{xy}$ symmetry
remains possible if the large gap effectively occurs on the
$\epsilon$ sheet. This conclusion is corroborated by recent
angle-resolved heat-capacity measurements of Kittaka {\it et
al}~\cite{Kittaka13}.
\begin{figure}[t]
\begin{center}
\includegraphics[width=9cm]{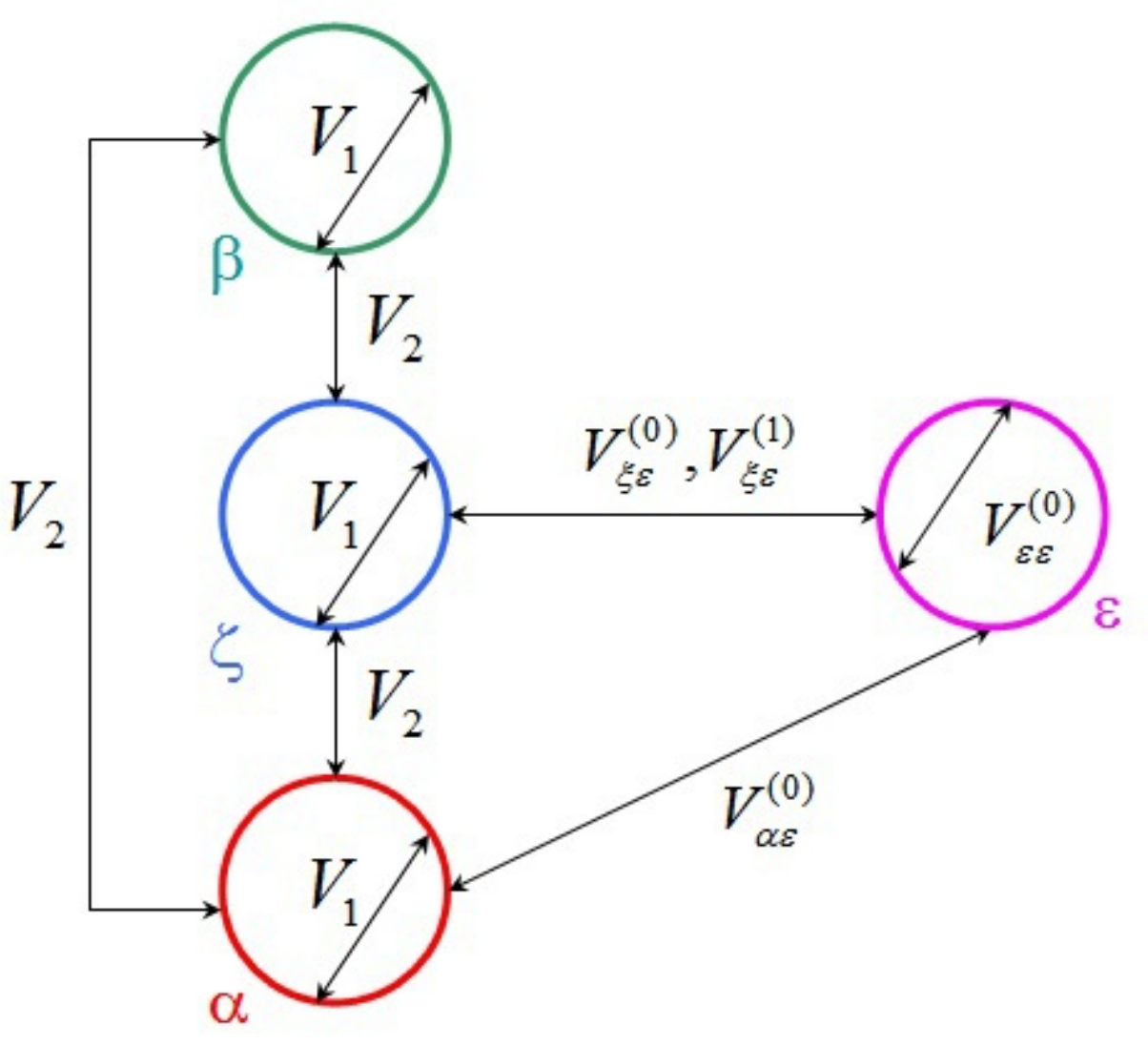}
\caption{\label{fig:Fig4A} (Color online) Schematics of the 4-band BCS model used to analyze C$_{e}$ (T, 0).}
\end{center}
\end{figure}

\section{Four-band BCS analysis of C$_{e}$ (T, 0)}
Recently, laser ARPES measurements,~\cite{Okazaki12} performed at
1.5 K, revealed angle-dependent gaps of the 3
bands around the $\Gamma$ point with accidental line nodes only on
the $\zeta$ sheet. These results are compatible with a nodal
$s$-state and exclude a possible change of symmetry of the
superconducting order parameter as a function of K doping. To check
whether these results can be confirmed by bulk measurements, we
analyze our zero-field heat capacity, taking into account the
observed Fermi surface. In the absence of a sizeable residual density of states for T $\rightarrow$ 0, the
modest specific-heat jump clearly shows that KFe$_{2}$As$_{2}$ is
close to the weak coupling limit. Therefore, we can model the
temperature dependence of C$_{e}$ in a pure 4-band BCS model.
Assuming 2D Fermi-surface pockets, we obtain the following system of
gap equations:
\begin{eqnarray}\label{eq:eq4}
&\Delta_{i}(\phi,T)=-\sum_{j=1}^{4} \frac{N_{j}(0)}{2 \pi}\int_{0}^{2 \pi}d\phi'\cdot\nonumber\\
&\int_{0}^{\epsilon_{c}}d\epsilon \frac{V_{ij}(\phi ,\phi')\Delta_{j}(\phi',T)}{\sqrt{\epsilon^{2}+|\Delta_{j}(\phi',T)|^{2}}}\tanh{\frac{\beta}{2}\sqrt{\epsilon^{2}+|\Delta_{j}(\phi',T)|^{2}}},
\end{eqnarray}
where N$_{i}$(0) is the density of states of the $i$-th band with $i\in \left\{\alpha,\zeta, \beta,\epsilon\right\}$, $V_{ij}(\phi, \phi')$ are the intraband ($i=j$) and interband ($i\neq j$) pairing potentials, $\beta$=1/k$_{B}$T, and $\phi$ and $\phi'$ the azimuthal angles on the sheets $i$ and $j$, respectively. In the $s$-wave channel,~\cite{Chubukov12,Maiti12} we write:
\begin{equation}\label{eq:eq3}
V_{ij}(\phi,\phi')=V_{ij}^{(0)}+V_{ij}^{(1)}\cdot(\cos{4\phi}+\cos{4\phi'}).
\end{equation}
Such interactions lead to anisotropic gaps of the form:
\begin{equation}\label{eq:eq5}
\Delta_{i}(\phi,T)=\Delta_{i}^{(0)}(T)+\Delta_{i}^{(1)}(T)\cdot\cos(4\phi),
\end{equation}
which are calculated self-consistently from Eqs. (\ref{eq:eq4}) and used to compute the superconducting-state heat capacity. We constrain all the N$_{i}$(0) to match as closely as possible the values inferred from dHvA and ARPES measurements. In this form, the model is parametrized with 20 interaction constants and this number is reduced to 5 by assuming that:
\begin{eqnarray}
V_{ij}^{(0)}&=&\delta_{ij}\cdot V_{1}+(1-\delta_{ij})\cdot V_{2},\label{eq:eq6a} \\
V_{ij}^{(1)}&=&0, \label{eq:eq6b} \\
V_{\beta \epsilon}(\phi,\phi')&=&V_{\epsilon \epsilon}^{(1)}=V_{\alpha \epsilon}^{(1)}=0,\label{eq:eq6c}
\end{eqnarray}
with $i,j\in \left\{\alpha,\zeta, \beta\right\}$. Here, Eqs. (\ref{eq:eq6a}) and (\ref{eq:eq6b}) impose that the intra- and interband interactions of the zone-centered bands are angle-independent and equal to V$_{1}$ and V$_{2}$, respectively because these bands have a quasi-2D morphology and are centered around the same point.~\cite{Maiti12} On the other hand, inelastic neutron scattering experiments have revealed the persistence of resonant spin excitations in heavily overdoped Ba$_{1-x}$K$_{1-x}$Fe$_{2}$As$_{2}$ (x $\approx$ 0.9)~\cite{Castellan11} and incommensurate spin fluctuations in KFe$_{2}$As$_{2}$ that approximately connect the $\Gamma$ and X bands.~\cite{Lee11} These observations convincingly indicate that the $\Gamma$-X interband interactions remain significant in KFe$_{2}$As$_{2}$, even in the absence of electron pockets.
\begin{figure*}[!t]
\includegraphics{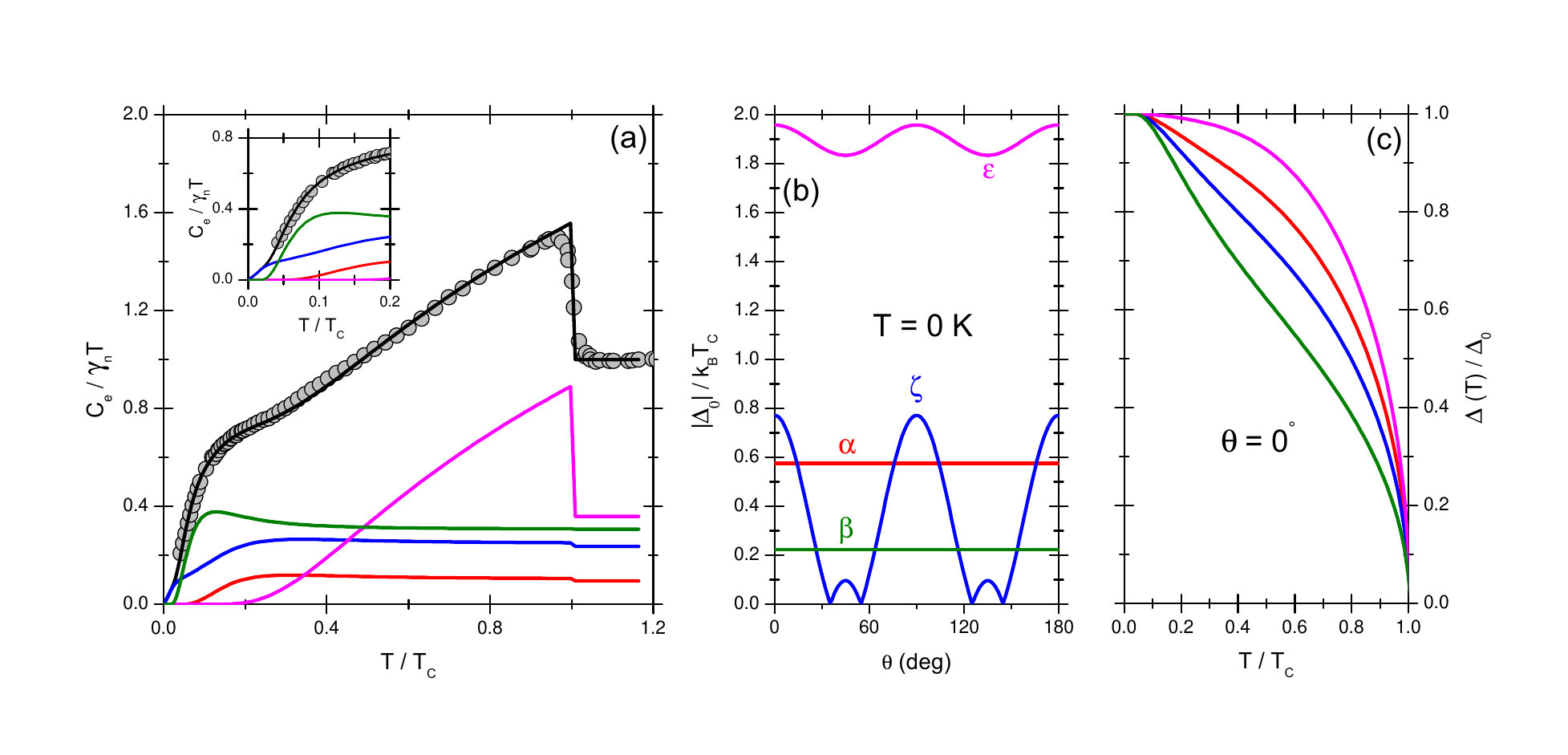}
\caption{\label{fig:Fig4} (Color online) (a) Temperature dependence
of the heat capacity of KFe$_{2}$As$_{2}$ derived in the 4-band BCS
model (black lines). Contributions from the individual bands are
also given. The inset shows T/T$_{c}$ $<$ 0.1 K. (b) Angular dependence of the individual gaps at T = 0 K. (c)
Temperature dependence of the indivdual gaps for $\theta$ = 0.}
\end{figure*}
This is particularly true for the $\alpha$ and $\zeta$ pockets which are strongly involved in nesting in Ba$_{0.6}$K$_{0.4}$Fe$_{2}$As$_{2}$. In KFe$_{2}$As$_{2}$, these bands share a dominant $xy/yz$ orbital character with the $\epsilon$ band while the $\beta$ pocket exhibits mainly $x^{2}-y^{2}$ component. This implies that the $\beta$ sheet plays no decisive role in pairing, as illustrated by the small gap observed on this pocket in both compounds.~\cite{Evtushinsky09,Okazaki12} Furthermore, the $\zeta$ band shows an additional finite $z^{2}$ component which is absent from the $\alpha$ and $\epsilon$ sheets. Consequently, the $z^{2}$ component in the $\zeta$ pocket has no counterpart for the sign change of the superconducting gap in the $\epsilon$ sheet and this can lead to a sign change of the gap in the parts of the $\zeta$ band where the $z^{2}$ contribution dominates.~\cite{Okazaki12} Thus, the only prominent angle-dependent interaction is V$_{\zeta \epsilon}^{(1)}$. All these assumptions are summarized in Eq. (\ref{eq:eq6c}) and Fig.\ref{fig:Fig4A}.

The heat capacity, as well as the angular and the temperature dependence of the gaps calculated with this model are shown in Fig.\ref{fig:Fig4}, using
the density of states and the remaining non-zero coupling constants
given in Tables \ref{tab:Table1} and \ref{tab:Table2}, respectively.
These parameters are all given in units of V$_{1}$ because it is not
their absolute values that matters, but rather their relative
weights. At the $\Gamma$ point, our results are in good agreement
with the laser ARPES experiments.
\begin{table}[h]
\caption{Fit parameters of the 4-band BCS model. The pairing
potentials are given in units of V$_{1}$ (with V$_{1}$=V$_{\alpha
\alpha}^{(0)}$=V$_{\zeta \zeta}^{(0)}$=V$_{\beta \beta}^{(0)}$ and
V$_{2}$=V$_{\alpha \zeta}^{(0)}$=V$_{\alpha \beta}^{(0)}$=V$_{\zeta
\beta}^{(0)}$). Comparison of the energy gaps derived from laser
ARPES (at 1.5 K), SANS (at 0.1 T), with those obtained from the
4-band BCS analysis of the zero-field heat capacity. The gaps are
given in units of k$_{B}$T$_{C}$. For the $\zeta$ and $\epsilon$
bands, the mean values are given. Assignments for the SANS data is
arbitrary.} \label{tab:Table2}
\begin{center}
\begin{tabular}{ccccc}
V$_{2}$     &   V$_{\epsilon \epsilon}^{(0)}$           &   V$_{\alpha \epsilon}^{(0)}$         &   V$_{\zeta \epsilon}^{(0)}$          &           V$_{\zeta \epsilon}^{(1)}$  \\
\hline
0.7             &   2.5                                                             & 0.5                                                       & 0.15                                                      &           0.6\\
\hline
\hline
                    &                                                                       &                                                                   &                                                                   &\\
\hline
\hline
\multicolumn{2}{c}{}                    &   ARPES                                                           &   SANS                                                        &   C (T)                                                   \\
\hline
\multicolumn{2}{c|}{$\alpha$}     & 3.8                                                             & 0.72                                                      & 0.57                                                          \\
\multicolumn{2}{c|}{$\beta$}        & 0.5                                                               & 0.21                                                      & 0.22                                                          \\
\multicolumn{2}{c|}{$\zeta$}        &   1.4                                                             & -                                                             & 0.35                                                          \\
\multicolumn{2}{c|}{$\epsilon$} &   -                                                                   & 1.77                                                      & 1.90
\end{tabular}
\end{center}
\end{table}
As shown in Fig. \ref{fig:Fig4}(b), the larger (smaller) of the 3 energy gaps is found on the $\alpha$ ($\beta$) band while the $\zeta$ gap exhibits accidental nodes which arise from the angle-dependent interband interaction with the $\epsilon$ pocket. These results are at odds with all theoretical calculations~\cite{Maiti11a,Maiti11b,Suzuki11,Thomale11} which predict the largest gap on the $\beta$ band. Moreover, at the X point, we recover the large gap $\Delta_{\epsilon}$=1.9 k$_{B}$T$_{c}$ inferred from our magnetization and field-dependent heat-capacity data. This gap was not observed in any other experiments and is due to a significantly larger intraband constant on the $\epsilon$ band (see Table \ref{tab:Table2}). We stress that our analysis is not unique and other sets of parameters could fit C$_{e}$(T) equally well. However, they would result necessarily in gap amplitudes close to the values we obtain because we constrain the individual densities of states to match approximately the values inferred from dHvA and ARPES. Moreover, as illustrated in Fig. \ref{fig:Fig4}(a), each gap has its own role in the temperature dependence of C$_{e}$(T) in the superconducting state. Indeed, the $\epsilon$ gap alone is responsible for the jump at T$_{c}$ and has a vanishing contribution at low temperatures, while the $\beta$ gap is predominantly responsible for the hump observed around T/T$_{c}$ $\approx$ 0.2. However, its decrease below this temperature is too steep to reproduce the experimental data. It is smoothened by the nodal contribution of the $\zeta$ gap. The slight maximum in the contributions from the $\alpha$ and $\zeta$ bands moreover lessen the dip of the high-temperature side of the shoulder due to the $\beta$ band. 

 Although our results do not bring direct evidence of the existence of line nodes, they firmly establish the existence of tiny energy gaps with $\Delta$/k$_{B}$T$_{c}$ $<$ 1.0. Their small amplitude imposes the requirement of cooling the sample below 80 mK to be able to observe the linear nodal behavior, as shown in the inset of Fig. \ref{fig:Fig4}(a). To our knowledge, no measurements in this temperature range were ever reported. Thus, our results do not exclude $d_{xy}$ symmetry. On the other hand, the agreement with laser ARPES is only qualitative. As shown in Table \ref{tab:Table2}, Okazaki {\it et al.} reported overestimated gap values in comparison to heat-capacity and SANS measurements. Particularly, they find $\Delta_{\alpha}$ = 3.8 k$_{B}$T$_{c}$, which is conspicuously comparable in amplitude to the largest gap observed close to the optimal concentration Ba$_{0.6}$K$_{0.4}$Fe$_{2}$As$_{2}$, while the critical temperature of the latter is 10 times larger.~\cite{Evtushinsky09,Xu11}\\

\section{Conclusions}
We have shown the existence of a relatively large nodeless energy
gap of amplitude 1.9 k$_{B}$T$_{c}$ that excludes the possibility of
$d_{x^{2}-y^{2}}$ symmetry for the superconducting-state order
parameter in KFe$_{2}$As$_{2}$. Our results do not bring direct
evidence for line nodes, they clearly prove the existence of tiny
energy gaps ($\Delta$/k$_{B}$T$_{c}$ $<$ 1.0) which strongly govern
the low-field and low-temperature heat capacity, much like MgB$_{2}$. Furthermore, the small
amplitudes of the gaps indicate that very low-temperature measurements (T
$<$ 80 mK) will be required in order to observe the possible signatures of line nodes in this
compound; a restriction that also applies to other probes like
penetration depth and heat transport. Our results shows qualitative
agreement with recent laser ARPES measurements, and strongly suggest
the superconducting-state symmetry to be $s$-wave.

\begin{acknowledgments}

\acknowledgments

We thank J.- P. Brison, M. Lang, M. Ichioka, K. Machida, A. Chubukov, T. Shibauchi and S. Kittaka for stimulating and enlightening discussions. This work was supported by the Deutsche Forschungsgemeinschaft through DFG-SPP 1458 "Hochtemperatursupraleitung in Eisenpniktiden". The work performed in Grenoble was supported by the French ANR Projects (SINUS and CHIRnMAG) and the ERC starting grant NewHeavyFermion.
\end{acknowledgments}





\end{document}